\documentclass[preprint1]{aastex62}
\accepted{\today}
\submitjournal{ApJ}

\shortauthors{Miao et al.}
\usepackage{url}
\usepackage{hyperref}
\newcommand{\nfig}[1]{Figure~\ref{#1}}
\newcommand{\speed}[1]{#1 km~s${}^{-1}$}

\newcommand{\sat}[1]{\it\uppercase{#1}\rm}
\newcommand{\kms}{km~s$^{-1}$}
\newcommand{\degree}{\ensuremath{^\circ}}

\begin{document}

\title{ A blowout jet associated with one obvious extreme-ultraviolet wave and one complicated coronal mass ejection event} 
\correspondingauthor{Y. Liu}
\email{lyu@ynao.ac.cn}
\correspondingauthor{Y. Miao}
\email{myh@ynao.ac.cn}
\author[0000-0003-2183-2095]{Y. Miao}
\affiliation{Yunnan Observatories, Chinese Academy of Sciences, Kunming, 650216, China}
\affiliation{Shandong Provincial Key Laboratory of Optical Astronomy and Solar-Terrestrial Environment，Shandong University, Weihai, 264209}
\affiliation{University of Chinese Academy of Sciences, Beijing 100049, China}
\author{Y. Liu}
\affiliation{Yunnan Observatories, Chinese Academy of Sciences, Kunming, 650216, China}
\author{H. B. Li}
\affiliation{Yunnan Observatories, Chinese Academy of Sciences, Kunming, 650216, China}
\affiliation{University of Chinese Academy of Sciences, Beijing 100049, China}
\author{Y. Shen}
\affiliation{Yunnan Observatories, Chinese Academy of Sciences, Kunming, 650216, China}
\affiliation{Center for Astronomical Mega-Science, Chinese Academy of Sciences, Beijing, 100012, China}
\affiliation{CAS Key Laboratory of Solar Activity, National Astronomical Observatories, Beijing 100012, China}
\author{S. Yang}
\affiliation{CAS Key Laboratory of Solar Activity, National Astronomical Observatories, Beijing 100012, China}
\author{A. ELMHAMDI}
\affiliation{Department of Physics and Astronomy, King Saud University, PO Box 2455, Riyadh 11451, Saudi Arabia}
\author{A. S. KORDI}
\affiliation{Department of Physics and Astronomy, King Saud University, PO Box 2455, Riyadh 11451, Saudi Arabia}
\author{Z. Z. ABIDIN}
\affiliation{Radio Cosmology Lab, Department of Physics, Faculty of Science, University of Malaya, 50603 Kuala Lumpur, Malaysia.}

\begin{abstract}
In this paper, we present a detailed analysis of a coronal blowout jet eruption which was associated with an obvious
extreme-ultraviolet (EUV) wave and one complicated coronal mass ejection (CME) event based on the multi-wavelength and
multi-view-angle observations from {\sl Solar Dynamics Observatory} and {\sl Solar Terrestrial Relations Observatory}.
It is found that the triggering of the blowout jet was due to the emergence and cancellation of magnetic fluxes on the
photosphere. During the rising stage of the jet, the EUV wave appeared just ahead of the jet top, lasting about 4
minutes and at a speed of 458 - \speed{762}. In addition, obvious dark material is observed along the EUV jet body,
which confirms the observation of a mini-filament eruption at the jet base in the chromosphere. Interestingly, two
distinct but overlapped CME structures can be observed in corona together with the eruption of the blowout jet. One is in narrow jet-shape, while the other one is in bubble-shape. The jet-shaped component was unambiguously related with the
outwardly running jet itself, while the bubble-like one might either be produced due to the reconstruction of the
high coronal fields or by the internal reconnection during the mini-filament ejection according to the double-CME
blowout jet model firstly proposed by Shen et al. (2012b), suggesting more observational evidence should be supplied to clear the current ambiguity based on large samples of blowout jets in future studies.
\end{abstract}

\section{Introduction}

Many kinds of jet activities have been observed in different layers of the solar
atmosphere with various spectral lines, such as the photosphere jets, chromospheric H$\alpha$ surges \citep{roy73,liu04}, low coronal extreme ultraviolet (EUV) and X-ray jets \citep{shib92,jiang07,li15,liu15,shen11}, and white-light jets observed by coronagraphs in the outer corona. They are thought to play an important role in explaining the problems of coronal heating and the acceleration of fast solar
wind \citep[e.g.,][]{shib92,jiang07,tian14,liu15}. \citet{shimo96,shimo98} analyzed X-ray jets and found important common characteristics. For example, the lifetime ranges from
a few minutes to more than ten hours; the length ranges from a few $\times$ 10$^{4}$ to 4 $\times$ 10$^{5}$ km; and the width ranges from 5 $\times$ 10$^{3}$ to 1 $\times$ 10$^{5}$ km; the velocity varies from \speed{10} to \speed{1000} and with an average velocity of about \speed{200}; 72\% of jets occur at the mixed polarity regions being mostly associated with small flares.

In the case of solar jets, however, the eruption properties are very different in dominant scales, velocities, configurations, and temperatures, while they have similar physical mechanism in nature. It is believed that jet eruptions are associated with magnetic flux emergence \citep{jiang07,2016SSRv..201....1R,shen17a}. For example, \cite{chae99} presented several active-region EUV jets that were thought to be caused by the magnetic cancellation between the newly emerging and the pre-existing opposite magnetic polarities. \citet{liu04} analyzed a surge that was driven by emerging magnetic fluxes associated with low reconnection on the photosphere. Magnetic reconnection
is thought to play an important role in the acceleration of solar jets \citep{shib94,shen11}. \citet{shib92} presented a standard two-dimensional
jet model, in which an emerging magnetic bipole consecutively presses against the ambient opposite polarity open fields and the magnetic
reconnection between open and closed fields that produces the jet. In recent years, this model has been widely tested in two-dimensional and
three-dimensional simulations and subsequently adopted by the scientific community \citep{yoko95,yoko96,pari09,arch10,moreno13,li17,tian17,zhu17}.

By using high resolution multi-wavelengths observations, \cite{moor10} found a new type of coronal jet that was dubbed ``blowout jets'' by the authors. For a typical blowout jet, it consists of a brighter base and a broader spire compared to the standard jet configuration. In the sample studied in \cite{moor10}, the authors found that about one-third of the jets are blowout jets, while the other two-third are standard jets. \cite{shen12} firstly reported the
existence of mini-filament eruption in the base of a blowout jet, and the authors augured that the formation of the cool component in the jet
was due to the eruption of the mini-filament, rather than the cooling of the hot plasma flow as previously thought
\citep[e.g.,][]{1994ApJ...425..326S,jiang07a}. Recently, \cite{shen17a} further confirmed their result by using high resolution observations
taken by the one-meter New Vacuum Solar Telescope. \cite{moor13} expanded upon their earlier work \citep{moor10} and found that about half
of the jets are indeed blowout jets. The authors used not only X-ray but also EUV observations in \cite{moor13}, and confirmed the results found
in \cite{shen12}, namely that almost all blowout jets were accompanied by the eruption of mini-filament (or cool material) in the jet base.
Many new studies further proposed that standard and blowout jets are fundamentally the same phenomenon, and all coronal jets are originated
from mini-filament eruptions \citep[e.g.,][]{shen12,2012ApJ...753..112Z,shen12b,shen17a,shen17b}. These investigations highly suggested the importance of mini-filament eruptions for coronal jets.

Based on the jet model proposed by \cite{moor10}, the dichotomy of coronal jets might be observed in the outer corona in coronagraphs.
Specifically, a standard jet will be observed as a single jet-like coronal mass eruptions (CMEs), while a blowout jet implies the production of
a jet-like and a simultaneous bubble-like CMEs. For standard jets, the corresponding CMEs in coronagraphs have been reported in various
previous studies \citep[e.g.,][]{1998ApJ...508..899W,2002ApJ...575..542W,2005ApJ...628.1056L,liu08,hong11,liu15}. However, so far the
observations of the corresponding simultaneous CMEs for coronal blowout jets is still very scarce. \cite{liu11} reported the on-disk
observations of a blowout jet that showed a loop-blob structure, and the authors conjectured that the erupting loop and blob might be
a miniature version of CMEs in the low corona as speculated by \cite{moor10}. The intriguing paired jet-like and bubble-like CMEs
evolving from a single blowout coronal jet was firstly reported by \cite{shen12} by using the stereoscopic high resolution observations
taken from three different view angles. The authors proposed that the jet-like CME was directly produced by the external magnetic
reconnection between open fields and the closed fields that confines the mini-filament, which is similar to the CMEs produced by
standard jets. The external reconnection removes the field overlying the mini-filament and therefore results in the rising and
eruption of the mini-filament, which subsequently evolves into the bubble-like CME in coronagraphs.

Global propagating extreme ultraviolet (EUV) waves were firstly detected by the Extreme-ultraviolet Imaging Telescope (EIT) onboard
the {\sl Solar and Heliospheric Observatory spacecraft} \citep[{\sl SOHO};][]{dela95,moses97,thomp98}. In the recent two decades, most
of the reported EUV waves were accompanied by large-scale energetic flares, and they were thought to be fast-mode magnetosonic waves
driven by CMEs \citep{2012ApJ...754....7S,2012ApJ...752L..23S,2013ApJ...773L..33S,shen14a,shen14b,li12,yang13}. In addition,
less energetic miniature eruptions can also lead to the appearance of EUV waves. Previous studies suggested that small-scale EUV waves
are often associated with miniature eruptions such as micro-sigmoid, mini-filament, and coronal jets
\citep[e.g.,][]{2012ApJ...753..112Z,zheng12,zheng13,2013MNRAS.431.1359Z,shen17b}. These miniature EUV waves have similar observational
characteristics with their large-scale counterparts, and they were also regarded as fast-mode magnetosonic waves in nature. Previous
studies indicated that micro EUV waves are often driven by expanding newly formed or pre-existing coronal loops caused by miniature
eruptions such as micro-sigmoid, mini-filament, and coronal jets \citep[e.g.,][]{2012ApJ...747...67Z,2012ApJ...753..112Z,zheng12}.
\cite{shen17b} recently reported the detailed formation process of a micro EUV wave associated with a mini-filament eruption, in
which the authors observed the separation of the EUV wave from the expanding coronal loop caused by the erupting mini-filament.
Although miniature eruptions do not cause CMEs in coronagraphs due to their insufficient energy nature, the eruption structures
including expanding loops and micro-sigmoids can be regarded as the initial state of CMEs in the low corona. Consequently, one can
consider this EUV wave as driven by CME, similar to large-scale EUV waves.

In this paper, we present the observational analysis of an intriguing coronal blowout jet that occurred on 2011 March 09 in NOAA active
region 11166, which was accompanied by a micro EUV wave and a paired simultaneous jet-like and bubble-like CMEs. This particular
event not only confirms the observational results previously presented in \cite{shen12} but also provides new clues for understanding the physics of coronal blowout jets. The paper is organized as follows: observations are introduced in Section 2; main observational results are
presented in Section 3; discussions and conclusions are given in Section 4.

\section{Observations}
The coronal blowout jet was simultaneously observed by {\em SDO}, {\em STEREO} Ahead (STA), and {\em STEREO} Behind (STB) from three
different view angles, and the positions of the three satellite are shown in \nfig{sat}. On 2011 March 09 at 22:00:00 UT, the separation
angle between STA and STB was about $177^\circ$, while that between STA (STB) and {\em SDO} was about $88^\circ$ ($95^\circ$). Since the
event was close to the disk center in the field-of-view (FOV) of {\em SDO}, it was on the west and east disk limb in the FOVs of STA and
STB, respectively. In addition, the eruption direction of the blowout jet is indicated by the green arrow in \nfig{sat}. The Helioseismic
and Magnetic Imager (HMI; \citealt{sche12}) onboard {\em SDO} provides full-disk line-of-sight (LOS) magnetograms with a pixel
width of $0\arcsec.5$ and a precision of 10 G with a 45 second cadence, while the Atmospheric Imaging Assembly (AIA; \citealt{lemen12})
onboard the {\sl SDO} has 12-second cadence and exposures of 0.12-2 seconds, and it images the Sun up to 1.3 R$_\sun$ in seven EUV and three
Ultraviolet (UV) wavelength bands with a pixel width of $0\arcsec.6$. The seven EUV wavelengths, 131 \AA, 171 \AA,
193 \AA, 211 \AA, 335 \AA, and 94 \AA, are due to strong iron lines with temperature range from 0.6 MK to 16 MK. The Extreme Ultraviolet
Imager (EUVI; \citealt{howard08}) of the Sun Earth Connection Coronal and Heliospheric Investigation (SECCHI; \citealt{howard08}) onboard
\sat{stereo} takes full-disk 304 \AA\ images, which have a 5 and 10 minute cadence and a pixel width of $1\arcsec.6$. The inner
coronagraphs (COR1) and the outer coronagraphs (COR2) (\citealt{thomp03}) onboard STA and STB take images with a cadence of 5 and 15 minutes,
and their FOVs are from 1.4 to 4 R$_\sun$ and 2.5 to 15.6 R$_\sun$, respectively.

\section{Results}
\subsection{The EUV Wave}
The eruption of the blowout jet was accompanied by a {\em GOES} C9.4 flare, whose start, peak, and end times were at 22:03:00, 22:12:00,
and 22:14:00 UT, respectively. The eruption of the jet started at about 22:05:00 UT and ended at about 22:25:00 UT, with hence a lifetime
of about 21 minutes. It is interesting that an arc-shaped EUV wave appeared on the southeastern side of the jet at about 22:06:34 UT,
which lasted for about 4 minutes. The EUV wave can be observed in the AIA 193, 171, and 211 \AA\ images. The propagating EUV wave is shown
in \nfig{wave} (a1) and (b1) with the AIA 171 and 193 \AA\  running difference images, and the wavefront has been highlighted by green dashed
curves. For detailed evolution process of the EUV wave, one can see the associated animation (animation1.mpeg) available in the online journal.
Some important parameters of the jet and the EUV wave are also listed in Table \ref{tab:list}. It is noted that the EUV wave only existed during
the earlier stage of the jet's evolution.

We use a semi-automatic method to obtain stack plots from ${20}^{\circ}$ wide sectors on the solar surface (see \citealt{liu10}).
In Figure \ref{slice_wave_position}, four sectors of A -- D are overlaid on the AIA 193 \AA\ full-disk running difference image
at 22:08:19 UT, and the position of the wavefront is also overlaid on the image as a dashed green curve. The stacked plots are made from
AIA 171, 193, and 211 \AA\ running difference images along the four sectors are plotted in Figure \ref{wave_velocity}. In each stack plot,
the propagating EUV wave can be observed as an inclined ridge. The slope of the ridge represents the velocity of the EUV wave, and it can be
obtained by fitting the ridge with a linear function. The measurement results show that the EUV wave had different velocities in different
directions. In stacked plots along Sector A (Figure \ref{wave_velocity} (a)--(c)), the wave signal only appeared in the 211 \AA\ plot, and the
wave velocity was about \speed{458}. In stacked plots along Sector B (Figure \ref{wave_velocity} (d)--(f)), the EUV wave appeared in the 193
and 211 \AA\ plots, which had a faster velocity of about \speed{625}. The EUV \textbf{wave} is best detected along Sectors C (Figure \ref{wave_velocity}
(g)--(i)) and D (Figure \ref{wave_velocity} (j)--(l)), and the velocities along the two directions were about  \speed{744} and \speed{762},
respectively.

\subsection{The simultaneous CME event}
The close up view of the jet's pre-eruption source region is shown in \nfig{filament} with the AIA 171 and 193 \AA\ images. A mini-filament can be
observed in the north-south direction, which was confined by a group of overlying loops. In \nfig{filament}, the profile of the mini-filament is
outlined with a green contour, while the overlying closed loops are indicated with dashed white curves. In the meantime, the details of the
mini-filament and the overlying loops are also magnified in the insets (see animation2.mpeg; available in the accompanying online material).

The evolution of the jet is shown in \nfig{aia} by using the AIA 171 \AA, 193 \AA, and 304 \AA\ observations. In the top row of \nfig{aia},
the mini-filament can be well identified in the insets which are running difference images of the box regions. The eruption of the jet started
at about 22:05:00 UT causing an obvious brightening of the source region (see the second row of \nfig{aia}). The eruption of the jet showed
clear untwisting motion, with the rotation direction indicated by the white arrow in \nfig{aia} (c2). According to
\citet{shib86,pari09,pari10,shen11}, the untwisting motion of jets can be attributed to the releasing of magnetic twist stored in the emerging closed bipole into the ambient open magnetic fields through magnetic magnetic reconnection. In addition, the rotation motion is also a typical
characteristic of coronal blowout jets \citep[e.g.,][]{shen12,moor13}. The violent eruption of the jet is displayed in the the bottom
two rows of \nfig{aia}. At about 22:18:33 UT, the jet body split into two parts, in which the preceding part continued to move outward, while the
following part started to fall back  (see the arrows in \nfig{wave} (a2) and (b2)). During the violent eruption phase of the jet, dark material
can be identified in the jet body (see the third row of \nfig{aia}). This indicates the eruption of the mini-filament confined by the closed
loops at the jet base region. Furthermore, it is noted that the brightening first occurred close to the south edge of the closed-loop system,
which suggests that the magnetic reconnection occurred between the closed-loop system and the ambient open field lines on the south of the
eruption source region.

The kinematics of the jet is studied with time-distance plots along and across the jet's main axis. The time-distance plots made from
AIA 171 \AA\ observations along the dashed lines (S1 and S2) in \nfig{slice} (a) are shown in \nfig{slice} (b) and (c), respectively.
The maximum length and width of the jet were measured about 435 and 40 Mm, respectively. The main ejection speed of the jet
was about \speed{350}, while the speeds of the preceding and following parts of the jet were about \speed{190 and -55}, respectively.

Figure \ref{extra} reports the results based on the force-field extrapolation (FFE) method \citep{zhu13,zhu16} applied to HMI vector magnetograms. Due to
the small size of the mini-filament, it is hard to determine this filament structure in the FFE model. However, the closed and the ambient open fields
can be well identified. We plot the profile of the mini-filament on the HMI LOS magentogram overlaid with the extrapolated field lines
and the AIA 171 \AA\ image at 22:02:00 UT (see \nfig{extra} (a) -- (c)), which indicate that the mini-filament laid on the neutral line
and was confined by the closed loops. Panels (d) and (e) show the extrapolated closed and the ambient open field lines as seen from different
viewpoints, with the magnetic polarities are also marked as ``P'', ``N'', and ``P1'', respectively. The extrapolated magnetic configuration
is well consistent with the observational results revealed by the AIA 171 \AA\ images as highlighted in \nfig{aia}.

In the top row of \nfig{hmi_euv}, the contours of the positive (red) and negative (blue) magnetic fields are overlaid on the
AIA 1600 and 1700 \AA\ images, in which the contour levels are $\pm100$ G, $\pm50$ G,  $\pm30$ G. The profile of the mini-filament
is also overlaid on the AIA 1600 and 1700 \AA\ images as a black contour, which indicates that the mini-filament laid on the magnetic
neutral line between polarities of P and N. The HMI LOS magnetograms are displayed in the bottom two rows of \nfig{hmi_euv}.

From (d) to (i), we use a series of HMI LOS magnetograms to emphasize the emergence and cancellation of the negative fluxes at the jet base. In panel (d), the green arrows indicate the cancellation of negative flux region and the red arrow indicates the emergence negative flux region, respectively. In order to indicate the detail of the cancellation, we plot three blue circles to display the very evident magnetic cancellation regions (see labels ``1'', ``2'', and ``3''). From (d) to (i), we can clearly recognize the negative flux cancellation and negative emergence evolution at the jet base region (JBR). From (d) to (i)
(from 21:55:59 UT to 22:55:59 UT), the region, indicated by a red arrow in panel (d), also shows the cancellation feature.
In panel (i), the three circles present very evident cancellation feature. This may indicate that magnetic cancellation and emergence
occurred at the JBR during the jet eruption in the whole stage and that the cancellation is more intense and quick than emergence. This is clear
in the online animation made from {\sl SDO}/HMI time sequence of images (animation4.mpeg).

In Figure \ref{flux_all}, the right panel is the magnified view of a sub-region highlighting the position of the jet base. The flux variation of the negative component, within the marked box, is reported in panel (c) from 21:31:14 UT to 22:38:44 UT. The maximum peak was at 21:55:59 UT, with a maximum reached absolute value of about $19.3$ $\times$ 10$^{19}$ Mx. The minimum was at around 22:32:44 UT, with a minimum absolute amount of about $17.6$ $\times$ 10$^{19}$ Mx.

It is important to pay more attention to the negative flux region because the positive flux region is too large and complex to precisely measure the positive flux evolution. In the box in panel (a) of Figure \ref{flux_all}, the negative flux is isolated, but the positive flux is not isolated (it crosses the boundary of the box). Therefore, in order to look for evidence of flux emergence or cancelation at the base of the jet around the time of jet onset, we track the changes with time of the negative flux with the box in Figure \ref{flux_all} (a). Figure \ref{flux_all} {b} shows a close-up of the box region, and panel(c) reports the changes with time of the negative flux in that box.
 From Figures \ref{hmi_euv} and \ref{flux_all}, the cancellation  during the jet eruption appears to be more intensive and rapid than during the emergence process within the time of interest. The magnetic cancellation can usually produce enough energy able to drive a large-scale jet eruption \citep{liu04}.

Interestingly, two different shaped CME structures, including a bubble-like and a jet-like structure, are observed to develop simultaneously with the top of the blowout jet that appeared in the FOVs of the COR1 Ahead and the COR1 Behind of the {\sl STEREO} mission. The separation angles from Earth was nearly 90$\degr$ for both {\sl STEREO A} and {B}, and the two satellites have successfully photographed the blowout jet eruption. Using the {\sl STEREO} EUVI 304 \AA\ and COR1 images, we can further investigate this CME event. From the two
viewpoints of {\sl STEREO} satellites, one can observe the intriguing blowout jet and its different characteristics.

The EUVI 304 \AA\ images and the COR1 running difference images are also used to show the evolution of the two simultaneous CME components in Figure \ref{stereo-cme}.
Panels (a1), (a2), (b1) and (b2) report the EUVI 304 \AA\ and COR1 combined images. The JBR and the jet are highlighted in panel (a1) in the FOV of the COR1 instrument. Panels (a3), (a4), (b3) and (b4) illustrate the
two CME components with running difference images. From this figure, the bubble-like component and the jet-like component seem to overlap symmetrically from the viewpoints of the two {\sl STEREO} spacecrafts, and they could be distinguished from time 22:35 UT to 22:55 UT in the FOVs of COR1 Ahead and Behind of {\sl STEREO}
(see animation5.mpeg). In panels (b3) and (b4), the red ``$\ast$'' mark denotes the top of the
jet-like component. It is noted that this top point seemed to surpass the bubble-like CME component at some time, such as at 22:45 (b4). The animation5
displays the whole stage of the double CME components evolution in FOVs of COR1. Especially, the bubble-like CME component seemed to catch up with the top of the jet-like component at about
22:55 UT. Then their apexes simultaneously escaped from the FOVs of COR1, and are hardly distinguishable afterwards. From Table \ref{tab:list}, the projected speed of the bubble-like component was faster than that of the jet-like one, averagely. Out of
the FOVs of COR1, the apex of the bubble-like component was measured as surpassing the jet-like component, because in the two FOVs of {\sl STEREO}/COR2 (an online movie highlights the process \footnote{
https://cdaw.gsfc.nasa.gov/movie/make$\_$javamovie.php?img1=stb$\_$cor2\&img2=sta$\_$cor2\&stime
=20110310$\_$0000\&etime=20110310$\_$0400\label{cor2}}) the bubble-like component was always ahead of the jet-like component before they disappeared from the field of view early on March 10. 
From the observations presented here, we can not decide whether the two CME components belong to a same CME.

\section{Discussion and Conclusions}

Solar jet activities have been studied widely with its direct relationship to large-scale coronal activities \citep[e.g.,][]{liu08,shen12,chen15,liu15,al16},
but the association of jet eruptions with EUV waves and CME events have been rarely reported. In this paper, we present one interesting blowout jet which can show close relation to both EUV wave and CME productions. During this blowout jet event, a C9.4 flare occurred. The evolution of the blowout jet has been analyzed in detail with the multi-angle observations from the {\sl STEREO}/EUVI/COR1/COR2 and {\sl SDO}/AIA/HMI instruments which had simultaneously captured the whole process of the eruption. In our jet event, some special observational characteristics are found. Firstly, the blowout jet was associated with an EUV wave running ahead at its top. Secondly, the blowout jet eruption was associated with
a mini-filament pre-existed at the base. Thirdly, clear CME evidence was found associated with the blowout jet, with one bubble-like CME front overlapped with one jet-like 'core' seen from the {\sl STEREO} LOS.

Additionally, from COR1 and COR2's FOVs we can also clearly observe the evolution of the two CME components. Especially, and according to Table \ref{tab:list}, the bubble-like component velocity was faster than the jet-like component after 22:45 UT. At about 04:00 UT on 10 March 2011, both the two CME components disappeared in the FOVs of COR2 A and B\textsuperscript{\ref{cor2}}.

At 22:25 UT, the jet preceding part became increasingly blurred out of the FOV of {\sl SDO}. We use then the {\sl STEREO} COR1 data to observe the two bright CME components. From the \sat{sdo} LOS, the jet projection on the disk of the Sun appears like a baseball bat. From the \sat{stereo} viewing angle, the jet-like CME component resembles a candle flame, whose top seems very close to the bubble-like structure. In Table \ref{tab:list}, the bubble-like and the jet-like CME components' velocities are listed to be \speed{220} and \speed{168}, respectively.

\citet{shib92} presented a standard X-ray jet model, and \citet{moor10} further compared the standard jet model and the blowout jet model. The blowout jet model is more complex than the standard jet model. \citet{moor10,moor15} pointed out that a standard jet could not possess enough free energy to produce a CME, contrary to the blowout jet events. On the other hand, due to the interior complex field at the blowout jet base, the highly sheared fields can drive more interchange reconnections which will cause the blowout jet to be heated to EUV and X-ray emission regions in a short time scale.

In the study of \citet{moor13}, a spinning motion in the eruptive growth phase was always observed for large X-ray jets. The cool component (mini-filament and other dark material
encompassed the closed-loop system) was easily observed from the {\sl SDO}/AIA 171, 193 and 304 \AA\ wavelengths. Similarly, in our study we also notice a discernible spinning motion from 22:05 UT to 22:15 UT based on the movie (animation3.mpeg) during the mini-filament
rising period. The blowout jet erupted with a C9.4 flare within the JBR. The eruption of the flare was so strong that a saturation effect was found within the data.

Our main findings are summarized as follows:

(1) In the jet base region, a bright arch patch was observed corresponding to magnetic flux cancellation process on the photosphere with an occurrence of a C9.4 flare.
The initiation of the blowout jet occurred at almost the same time as the start of the brightening. These phenomena are explained as the manifestations of the external reconnection of the blowout jet in the lower atmosphere, while the reconnection site was possibly located under the upper chromosphere.

(2) On the one hand, the wave front expansion speed ranges from \speed{458} to \speed{762} in different directions, roughly of the order of the average surface projected expansion speeds for fast-mode waves \citep{wang00}. These speeds appear greater than typical EUV waves speeds of \speed{200-400}\citep{klass00,thomp09}. The EUV wave was faster than the blowout jet speed ({\speed{190-350}), and the EUV wave front was easily distinguishable from the bubble-like CME structure. On the other hand, the wave was possibly triggered by the blowout jet or by the flare, but we favor the explanation that it was triggered by the blowout jet due to their close timing and location relations. The wave departed far from the flare center and showed a close location relative to the rapid blowout jet (animation1.mpeg). All the findings provide the evidence that the EUV wave was a fast-mode MHD wave and driven by the blowout jet.

(3) The blowout jet displayed both cool and hot components inside. The cool component (see Figure \ref{aia} dark material) was identified from the erupting mini-filament, while the hot component was considered as the hot
outflow generated by the external reconnection. This motion was observed at the beginning of the erupting blowout jet. The untwisting motion can be interpreted by using the reconnection models that use induced magnetic unwinding as the driving mechanism \citep{shib86,pari09,pari10}. The average speed of the jet was about \speed{350}. At about 22:18 UT, the blowout jet split into two parts, and the speeds of the preceding and following parts of the jet were about \speed{190 and -55}, respectively.

(4) From the {\sl STEREO}/Ahead and Behind two viewpoints, a simultaneous bubble-like and a jet-like CME components
appeared in the FOVs of COR1 and COR2. We find that the jet-like component was the extension of the hot component
of the blowout jet body in the outer corona, while the bubble-like component was
associated with the eruption of a mini-filament at the jet base. Physically, we suggested that the bubble-like front
was produced either due to the reconstruction of the high coronal fields or by the internal reconnection during the
mini-filament ejection according to the double-CME blowout jet model by \citet{shen12}. From Table \ref{tab:list}, the bubble-like CME component velocity is faster than that of the jet-like component.

 Worth to note here that this blowout jet eruption may present another case to support the scenario argued by \citet{shen12} and \citet{zheng13}, in which an EUV wave and two different CME components (one bubble-like and one jet-like) can be produced and associated with each other in the blowout jet model. In other words, simultaneous bubble-like CME and jet-like CME may be related to the same coronal blowout jet. Nonetheless, the association phenomenon of a blowout jet with either EUV wave or CME event is still rather rare and unclear. We need more similar investigations to figure out the key process of the blowout jets in the future studies.

\acknowledgments We thank the excellent data provided by the {\em SDO} and {\em STEREO} teams.
 We also thank the referee for his/her valuable suggestions and comments that improved the quality of the paper. This work is funded by the grants from the National Scientific Foundation of China (NSFC 11533009, 11773068), and the Project Supported by the Specialized Research Fund for Shandong Provincial Key Laboratory. This work is also supported by the grant associated with project of the Group for Innovation of Yunnan Province
and the Strategic Priority Research Program of CAS with grant XDA-17040507. The authors Y. Liu and Z. Z. Abidin would like to thank the University of Malaya Faculty of Science Grant (GPF040B-2018) for their support. The research by A. Elmhamdi was supported by King Saud University, Deanship of Scientific Research, College of Science Research Center. In addition, we are also grateful to the One Belt and One Road project of the West Light Foundation, CAS.


\begin{thebibliography}{}
\expandafter\ifx\csname natexlab\endcsname\relax\def\natexlab#1{#1}\fi

\bibitem[{{Alzate} \& {Morgan}(2016)}]{al16}
{Alzate}, N., \& {Morgan}, H. 2016, \apj, 823, 129

\bibitem[{{Archontis} \& {Hood}(2010)}]{arch10}
{Archontis}, V., \& {Hood}, A.~W. 2010, \aap, 514, A56

\bibitem[{{Chae} {et~al.}(1999){Chae}, {Qiu}, {Wang}, \& {Goode}}]{chae99}
{Chae}, J., {Qiu}, J., {Wang}, H., \& {Goode}, P.~R. 1999, \apjl, 513, L75

\bibitem[{{Chen} {et~al.}(2015){Chen}, {Su}, {Yin}, {Priya}, {Zhang}, {Liu},
  {Xu}, \& {Yu}}]{chen15}
{Chen}, J., {Su}, J., {Yin}, Z., {et~al.} 2015, \apj, 815, 71

\bibitem[{{Delaboudini{\`e}re} {et~al.}(1995){Delaboudini{\`e}re}, {Artzner},
  {Brunaud}, {Gabriel}, {Hochedez}, {Millier}, {Song}, {Au}, {Dere}, {Howard},
  {Kreplin}, {Michels}, {Moses}, {Defise}, {Jamar}, {Rochus}, {Chauvineau},
  {Marioge}, {Catura}, {Lemen}, {Shing}, {Stern}, {Gurman}, {Neupert},
  {Maucherat}, {Clette}, {Cugnon}, \& {van Dessel}}]{dela95}
{Delaboudini{\`e}re}, J.-P., {Artzner}, G.~E., {Brunaud}, J., {et~al.} 1995,
  \solphys, 162, 291

\bibitem[{{Hong} {et~al.}(2011){Hong}, {Jiang}, {Zheng}, {Yang}, {Bi}, \&
  {Yang}}]{hong11}
{Hong}, J., {Jiang}, Y., {Zheng}, R., {et~al.} 2011, \apjl, 738, L20

\bibitem[{{Howard} {et~al.}(2008){Howard}, {Moses}, {Vourlidas}, {Newmark},
  {Socker}, {Plunkett}, {Korendyke}, {Cook}, {Hurley}, {Davila}, {Thompson},
  {St Cyr}, {Mentzell}, {Mehalick}, {Lemen}, {Wuelser}, {Duncan}, {Tarbell},
  {Wolfson}, {Moore}, {Harrison}, {Waltham}, {Lang}, {Davis}, {Eyles},
  {Mapson-Menard}, {Simnett}, {Halain}, {Defise}, {Mazy}, {Rochus}, {Mercier},
  {Ravet}, {Delmotte}, {Auchere}, {Delaboudiniere}, {Bothmer}, {Deutsch},
  {Wang}, {Rich}, {Cooper}, {Stephens}, {Maahs}, {Baugh}, {McMullin}, \&
  {Carter}}]{howard08}
{Howard}, R.~A., {Moses}, J.~D., {Vourlidas}, A., {et~al.} 2008, \ssr, 136, 67

\bibitem[{{Jiang} {et~al.}(2007{\natexlab{a}}){Jiang}, {Chen}, {Shen}, {Yang},
  \& {Li}}]{jiang07a}
{Jiang}, Y., {Chen}, H., {Shen}, Y., {Yang}, L., \& {Li}, K.
  2007{\natexlab{a}}, \solphys, 240, 77

\bibitem[{{Jiang} {et~al.}(2007{\natexlab{b}}){Jiang}, {Chen}, {Li}, {Shen}, \&
  {Yang}}]{jiang07}
{Jiang}, Y.~C., {Chen}, H.~D., {Li}, K.~J., {Shen}, Y.~D., \& {Yang}, L.~H.
  2007{\natexlab{b}}, \aap, 469, 331

\bibitem[{{Klassen} {et~al.}(2000){Klassen}, {Aurass}, {Mann}, \&
  {Thompson}}]{klass00}
{Klassen}, A., {Aurass}, H., {Mann}, G., \& {Thompson}, B.~J. 2000, \aaps, 141,
  357

\bibitem[{{Lemen} {et~al.}(2012){Lemen}, {Title}, {Akin}, {Boerner}, {Chou},
  {Drake}, {Duncan}, {Edwards}, {Friedlaender}, {Heyman}, {Hurlburt}, {Katz},
  {Kushner}, {Levay}, {Lindgren}, {Mathur}, {McFeaters}, {Mitchell}, {Rehse},
  {Schrijver}, {Springer}, {Stern}, {Tarbell}, {Wuelser}, {Wolfson}, {Yanari},
  {Bookbinder}, {Cheimets}, {Caldwell}, {Deluca}, {Gates}, {Golub}, {Park},
  {Podgorski}, {Bush}, {Scherrer}, {Gummin}, {Smith}, {Auker}, {Jerram},
  {Pool}, {Soufli}, {Windt}, {Beardsley}, {Clapp}, {Lang}, \&
  {Waltham}}]{lemen12}
{Lemen}, J.~R., {Title}, A.~M., {Akin}, D.~J., {et~al.} 2012, \solphys, 275, 17

\bibitem[{{Li} {et~al.}(2017){Li}, {Jiang}, {Yang}, {Yang}, {Xu}, {Hong}, \&
  {Bi}}]{li17}
{Li}, H., {Jiang}, Y., {Yang}, J., {et~al.} 2017, \apj, 836, 235

\bibitem[{{Li} \& {Zhang}(2012)}]{li12}
{Li}, T., \& {Zhang}, J. 2012, \apjl, 760, L10

\bibitem[{{Li} {et~al.}(2015){Li}, {Yang}, {Chen}, {Li}, \& {Zhang}}]{li15}
{Li}, X., {Yang}, S., {Chen}, H., {Li}, T., \& {Zhang}, J. 2015, \apjl, 814,
  L13

\bibitem[{{Liu} {et~al.}(2011){Liu}, {Deng}, {Liu}, {Ugarte-Urra}, {Wang}, \&
  {Wang}}]{liu11}
{Liu}, C., {Deng}, N., {Liu}, R., {et~al.} 2011, \apjl, 735, L18

\bibitem[{{Liu} {et~al.}(2015){Liu}, {Wang}, {Shen}, {Liu}, {Pan}, \&
  {Wang}}]{liu15}
{Liu}, J., {Wang}, Y., {Shen}, C., {et~al.} 2015, \apj, 813, 115

\bibitem[{{Liu} {et~al.}(2010){Liu}, {Nitta}, {Schrijver}, {Title}, \&
  {Tarbell}}]{liu10}
{Liu}, W., {Nitta}, N.~V., {Schrijver}, C.~J., {Title}, A.~M., \& {Tarbell},
  T.~D. 2010, \apjl, 723, L53

\bibitem[{{Liu}(2008)}]{liu08}
{Liu}, Y. 2008, \solphys, 249, 75

\bibitem[{{Liu} \& {Kurokawa}(2004)}]{liu04}
{Liu}, Y., \& {Kurokawa}, H. 2004, \apj, 610, 1136

\bibitem[{{Liu} {et~al.}(2005){Liu}, {Su}, {Morimoto}, {Kurokawa}, \&
  {Shibata}}]{2005ApJ...628.1056L}
{Liu}, Y., {Su}, J.~T., {Morimoto}, T., {Kurokawa}, H., \& {Shibata}, K. 2005,
  \apj, 628, 1056

\bibitem[{{Moore} {et~al.}(2010){Moore}, {Cirtain}, {Sterling}, \&
  {Falconer}}]{moor10}
{Moore}, R.~L., {Cirtain}, J.~W., {Sterling}, A.~C., \& {Falconer}, D.~A. 2010,
  \apj, 720, 757

\bibitem[{{Moore} {et~al.}(2013){Moore}, {Sterling}, {Falconer}, \&
  {Robe}}]{moor13}
{Moore}, R.~L., {Sterling}, A.~C., {Falconer}, D.~A., \& {Robe}, D. 2013, \apj,
  769, 134

\bibitem[{{Moore} {et~al.}(2015){Moore}, {Sterling}, \& {Falconer}}]{moor15}
{Moore}, R.~L., {Sterling}, A.~C., \& {Falconer}, D.~A. 2015, \apj, 806, 11

\bibitem[{{Moreno-Insertis} \& {Galsgaard}(2013)}]{moreno13}
{Moreno-Insertis}, F., \& {Galsgaard}, K. 2013, \apj, 771, 20

\bibitem[{{Moses} {et~al.}(1997){Moses}, {Clette}, {Delaboudini{\`e}re},
  {Artzner}, {Bougnet}, {Brunaud}, {Carabetian}, {Gabriel}, {Hochedez},
  {Millier}, {Song}, {Au}, {Dere}, {Howard}, {Kreplin}, {Michels}, {Defise},
  {Jamar}, {Rochus}, {Chauvineau}, {Marioge}, {Catura}, {Lemen}, {Shing},
  {Stern}, {Gurman}, {Neupert}, {Newmark}, {Thompson}, {Maucherat},
  {Portier-Fozzani}, {Berghmans}, {Cugnon}, {van Dessel}, \&
  {Gabryl}}]{moses97}
{Moses}, D., {Clette}, F., {Delaboudini{\`e}re}, J.-P., {et~al.} 1997,
  \solphys, 175, 571

\bibitem[{{Pariat} {et~al.}(2009){Pariat}, {Antiochos}, \& {DeVore}}]{pari09}
{Pariat}, E., {Antiochos}, S.~K., \& {DeVore}, C.~R. 2009, \apj, 691, 61

\bibitem[{{Pariat} {et~al.}(2010){Pariat}, {Antiochos}, \& {DeVore}}]{pari10}
---. 2010, \apj, 714, 1762

\bibitem[{{Raouafi} {et~al.}(2016){Raouafi}, {Patsourakos}, {Pariat}, {Young},
  {Sterling}, {Savcheva}, {Shimojo}, {Moreno-Insertis}, {DeVore}, {Archontis},
  {T{\"o}r{\"o}k}, {Mason}, {Curdt}, {Meyer}, {Dalmasse}, \&
  {Matsui}}]{2016SSRv..201....1R}
{Raouafi}, N.~E., {Patsourakos}, S., {Pariat}, E., {et~al.} 2016, \ssr, 201, 1

\bibitem[{{Roy}(1973)}]{roy73}
{Roy}, J.~R. 1973, \solphys, 28, 95

\bibitem[{{Scherrer} {et~al.}(2012){Scherrer}, {Schou}, {Bush}, {Kosovichev},
  {Bogart}, {Hoeksema}, {Liu}, {Duvall}, {Zhao}, {Title}, {Schrijver},
  {Tarbell}, \& {Tomczyk}}]{sche12}
{Scherrer}, P.~H., {Schou}, J., {Bush}, R.~I., {et~al.} 2012, \solphys, 275,
  207

\bibitem[{{Schmieder} {et~al.}(1994){Schmieder}, {Golub}, \&
  {Antiochos}}]{1994ApJ...425..326S}
{Schmieder}, B., {Golub}, L., \& {Antiochos}, S.~K. 1994, \apj, 425, 326

\bibitem[{{Shen} {et~al.}(2014{\natexlab{a}}){Shen}, {Ichimoto}, {Ishii},
  {Tian}, {Zhao}, \& {Shibata}}]{shen14a}
{Shen}, Y., {Ichimoto}, K., {Ishii}, T.~T., {et~al.} 2014{\natexlab{a}}, \apj,
  786, 151

\bibitem[{{Shen} \& {Liu}(2012{\natexlab{a}})}]{2012ApJ...754....7S}
{Shen}, Y., \& {Liu}, Y. 2012{\natexlab{a}}, \apj, 754, 7

\bibitem[{{Shen} \& {Liu}(2012{\natexlab{b}})}]{2012ApJ...752L..23S}
---. 2012{\natexlab{b}}, \apjl, 752, L23

\bibitem[{{Shen} {et~al.}(2012{\natexlab{a}}){Shen}, {Liu}, \& {Su}}]{shen12b}
{Shen}, Y., {Liu}, Y., \& {Su}, J. 2012{\natexlab{a}}, \apj, 750, 12

\bibitem[{{Shen} {et~al.}(2012{\natexlab{b}}){Shen}, {Liu}, {Su}, \&
  {Deng}}]{shen12}
{Shen}, Y., {Liu}, Y., {Su}, J., \& {Deng}, Y. 2012{\natexlab{b}}, \apj, 745,
  164

\bibitem[{{Shen} {et~al.}(2011){Shen}, {Liu}, {Su}, \& {Ibrahim}}]{shen11}
{Shen}, Y., {Liu}, Y., {Su}, J., \& {Ibrahim}, A. 2011, \apjl, 735, L43

\bibitem[{{Shen} {et~al.}(2013){Shen}, {Liu}, {Su}, {Li}, {Zhao}, {Tian},
  {Ichimoto}, \& {Shibata}}]{2013ApJ...773L..33S}
{Shen}, Y., {Liu}, Y., {Su}, J., {et~al.} 2013, \apjl, 773, L33

\bibitem[{{Shen} {et~al.}(2017{\natexlab{a}}){Shen}, {Liu}, {Tian}, \&
  {Qu}}]{shen17b}
{Shen}, Y., {Liu}, Y., {Tian}, Z., \& {Qu}, Z. 2017{\natexlab{a}}, \apj, 851,
  101

\bibitem[{{Shen} {et~al.}(2014{\natexlab{b}}){Shen}, {Liu}, {Chen}, \&
  {Ichimoto}}]{shen14b}
{Shen}, Y., {Liu}, Y.~D., {Chen}, P.~F., \& {Ichimoto}, K. 2014{\natexlab{b}},
  \apj, 795, 130

\bibitem[{{Shen} {et~al.}(2017{\natexlab{b}}){Shen}, {Liu}, {Su}, {Qu}, \&
  {Tian}}]{shen17a}
{Shen}, Y., {Liu}, Y.~D., {Su}, J., {Qu}, Z., \& {Tian}, Z. 2017{\natexlab{b}},
  \apj, 851, 67

\bibitem[{{Shibata} {et~al.}(1994){Shibata}, {Nitta}, {Strong}, {Matsumoto},
  {Yokoyama}, {Hirayama}, {Hudson}, \& {Ogawara}}]{shib94}
{Shibata}, K., {Nitta}, N., {Strong}, K.~T., {et~al.} 1994, \apjl, 431, L51

\bibitem[{{Shibata} \& {Uchida}(1986)}]{shib86}
{Shibata}, K., \& {Uchida}, Y. 1986, \solphys, 103, 299

\bibitem[{{Shibata} {et~al.}(1992){Shibata}, {Ishido}, {Acton}, {Strong},
  {Hirayama}, {Uchida}, {McAllister}, {Matsumoto}, {Tsuneta}, {Shimizu},
  {Hara}, {Sakurai}, {Ichimoto}, {Nishino}, \& {Ogawara}}]{shib92}
{Shibata}, K., {Ishido}, Y., {Acton}, L.~W., {et~al.} 1992, \pasj, 44, L173

\bibitem[{{Shimojo} {et~al.}(1996){Shimojo}, {Hashimoto}, {Shibata},
  {Hirayama}, {Hudson}, \& {Acton}}]{shimo96}
{Shimojo}, M., {Hashimoto}, S., {Shibata}, K., {et~al.} 1996, \pasj, 48, 123

\bibitem[{{Shimojo} {et~al.}(1998){Shimojo}, {Shibata}, {Yokoyama}, \&
  {Hori}}]{shimo98}
{Shimojo}, M., {Shibata}, K., {Yokoyama}, T., \& {Hori}, K. 1998, in ESA
  Special Publication, Vol. 421, Solar Jets and Coronal Plumes, ed. T.-D.
  {Guyenne}, 163

\bibitem[{{Thompson} \& {Myers}(2009)}]{thomp09}
{Thompson}, B.~J., \& {Myers}, D.~C. 2009, \apjs, 183, 225

\bibitem[{{Thompson} {et~al.}(1998){Thompson}, {Plunkett}, {Gurman}, {Newmark},
  {St.~Cyr}, \& {Michels}}]{thomp98}
{Thompson}, B.~J., {Plunkett}, S.~P., {Gurman}, J.~B., {et~al.} 1998, \grl, 25,
  2465

\bibitem[{{Thompson} {et~al.}(2003){Thompson}, {Davila}, {Fisher}, {Orwig},
  {Mentzell}, {Hetherington}, {Derro}, {Federline}, {Clark}, {Chen},
  {Tveekrem}, {Martino}, {Novello}, {Wesenberg}, {StCyr}, {Reginald}, {Howard},
  {Mehalick}, {Hersh}, {Newman}, {Thomas}, {Card}, \& {Elmore}}]{thomp03}
{Thompson}, W.~T., {Davila}, J.~M., {Fisher}, R.~R., {et~al.} 2003, in
  \procspie, Vol. 4853, Innovative Telescopes and Instrumentation for Solar
  Astrophysics, ed. S.~L. {Keil} \& S.~V. {Avakyan}, 1--11

\bibitem[{{Tian} {et~al.}(2014){Tian}, {DeLuca}, {Cranmer}, {De Pontieu},
  {Peter}, {Mart{\'{\i}}nez-Sykora}, {Golub}, {McKillop}, {Reeves}, {Miralles},
  {McCauley}, {Saar}, {Testa}, {Weber}, {Murphy}, {Lemen}, {Title}, {Boerner},
  {Hurlburt}, {Tarbell}, {Wuelser}, {Kleint}, {Kankelborg}, {Jaeggli},
  {Carlsson}, {Hansteen}, \& {McIntosh}}]{tian14}
{Tian}, H., {DeLuca}, E.~E., {Cranmer}, S.~R., {et~al.} 2014, Science, 346,
  1255711

\bibitem[{{Tian} {et~al.}(2017){Tian}, {Liu}, {Shen}, {Elmhamdi}, {Su}, {Liu},
  \& {Kordi}}]{tian17}
{Tian}, Z., {Liu}, Y., {Shen}, Y., {et~al.} 2017, \apj, 845, 94

\bibitem[{{Wang}(2000)}]{wang00}
{Wang}, Y.-M. 2000, \apjl, 543, L89

\bibitem[{{Wang} \& {Sheeley}(2002)}]{2002ApJ...575..542W}
{Wang}, Y.-M., \& {Sheeley}, Jr., N.~R. 2002, \apj, 575, 542

\bibitem[{{Wang} {et~al.}(1998){Wang}, {Sheeley}, {Socker}, {Howard},
  {Brueckner}, {Michels}, {Moses}, {St.~Cyr}, {Llebaria}, \&
  {Delaboudini{\`e}re}}]{1998ApJ...508..899W}
{Wang}, Y.-M., {Sheeley}, Jr., N.~R., {Socker}, D.~G., {et~al.} 1998, \apj,
  508, 899

\bibitem[{{Yang} {et~al.}(2013){Yang}, {Zhang}, {Liu}, {Li}, \&
  {Shen}}]{yang13}
{Yang}, L., {Zhang}, J., {Liu}, W., {Li}, T., \& {Shen}, Y. 2013, \apj, 775, 39

\bibitem[{{Yokoyama} \& {Shibata}(1995)}]{yoko95}
{Yokoyama}, T., \& {Shibata}, K. 1995, \nat, 375, 42

\bibitem[{{Yokoyama} \& {Shibata}(1996)}]{yoko96}
---. 1996, \pasj, 48, 353

\bibitem[{{Zheng} {et~al.}(2012{\natexlab{a}}){Zheng}, {Jiang}, {Yang}, {Bi},
  {Hong}, {Yang}, \& {Yang}}]{2012ApJ...747...67Z}
{Zheng}, R., {Jiang}, Y., {Yang}, J., {et~al.} 2012{\natexlab{a}}, \apj, 747,
  67

\bibitem[{{Zheng} {et~al.}(2013{\natexlab{a}}){Zheng}, {Jiang}, {Yang}, {Bi},
  {Hong}, {Yang}, \& {Yang}}]{zheng13}
---. 2013{\natexlab{a}}, \apj, 764, 70

\bibitem[{{Zheng} {et~al.}(2012{\natexlab{b}}){Zheng}, {Jiang}, {Yang}, {Bi},
  {Hong}, {Yang}, \& {Yang}}]{2012ApJ...753..112Z}
---. 2012{\natexlab{b}}, \apj, 753, 112

\bibitem[{{Zheng} {et~al.}(2012{\natexlab{c}}){Zheng}, {Jiang}, {Yang}, {Bi},
  {Hong}, {Yang}, \& {Yang}}]{zheng12}
---. 2012{\natexlab{c}}, \apjl, 753, L29

\bibitem[{{Zheng} {et~al.}(2013{\natexlab{b}}){Zheng}, {Jiang}, {Yang}, {Hong},
  {Bi}, {Yang}, \& {Yang}}]{2013MNRAS.431.1359Z}
{Zheng}, R.-S., {Jiang}, Y.-C., {Yang}, J.-Y., {et~al.} 2013{\natexlab{b}},
  \mnras, 431, 1359

\bibitem[{{Zhu} {et~al.}(2017){Zhu}, {Wang}, {Cheng}, \& {Huang}}]{zhu17}
{Zhu}, X., {Wang}, H., {Cheng}, X., \& {Huang}, C. 2017, \apjl, 844, L20

\bibitem[{{Zhu} {et~al.}(2016){Zhu}, {Wang}, {Du}, \& {He}}]{zhu16}
{Zhu}, X., {Wang}, H., {Du}, Z., \& {He}, H. 2016, \apj, 826, 51

\bibitem[{{Zhu} {et~al.}(2013){Zhu}, {Wang}, {Du}, \& {Fan}}]{zhu13}
{Zhu}, X.~S., {Wang}, H.~N., {Du}, Z.~L., \& {Fan}, Y.~L. 2013, \apj, 768, 119

\end{thebibliography}

\begin{table*}[]
\begin{center}
\caption{Parameters of the observed blowout jet, the EUV wave, and the CME components. \label{tab:list}}
\scalebox{1}{
\begin{tabular}{c*{8}{c}}
  \noalign{\smallskip}\tableline\tableline \noalign{\smallskip}
  \multicolumn{8}{l}{} \\
   \noalign{\smallskip} \noalign{\smallskip}

    Event & Start Time & Flare\footnote{https://www.solarmonitor.org/?date=20110309} & Speed  & Angular       & End Time & Dur. Time \\
 2011-Mar-09 & (UT)     & class  & (\kms)                 & inclination (\degree) & (UT)   & (min)  \\
\noalign{\smallskip}\tableline \noalign{\smallskip}

Jet & 22:05:00  & C9.4   & 190/350\footnote{the main ejection speed of the jet
is about \speed{350} and the preceding part average velocity is about \speed{190}, respectively.}  & 40$\pm$ 5  & 22:25:00  & 20 & \\
Wave & 22:06:34 & \nodata   & 458-762 & \nodata  & 22:10:34     & 4  \\
CME components & 22:25:00  & \nodata    & 220/168\footnote{the bubble-like CME component average velocity is about \speed{220} and the jet-like is about \speed{168}} & \nodata     & 04:00:00+1 day    &  355 \\

\tableline
\end{tabular}}

\end{center}
\end{table*}

\begin{figure}
\epsscale{1} \plotone{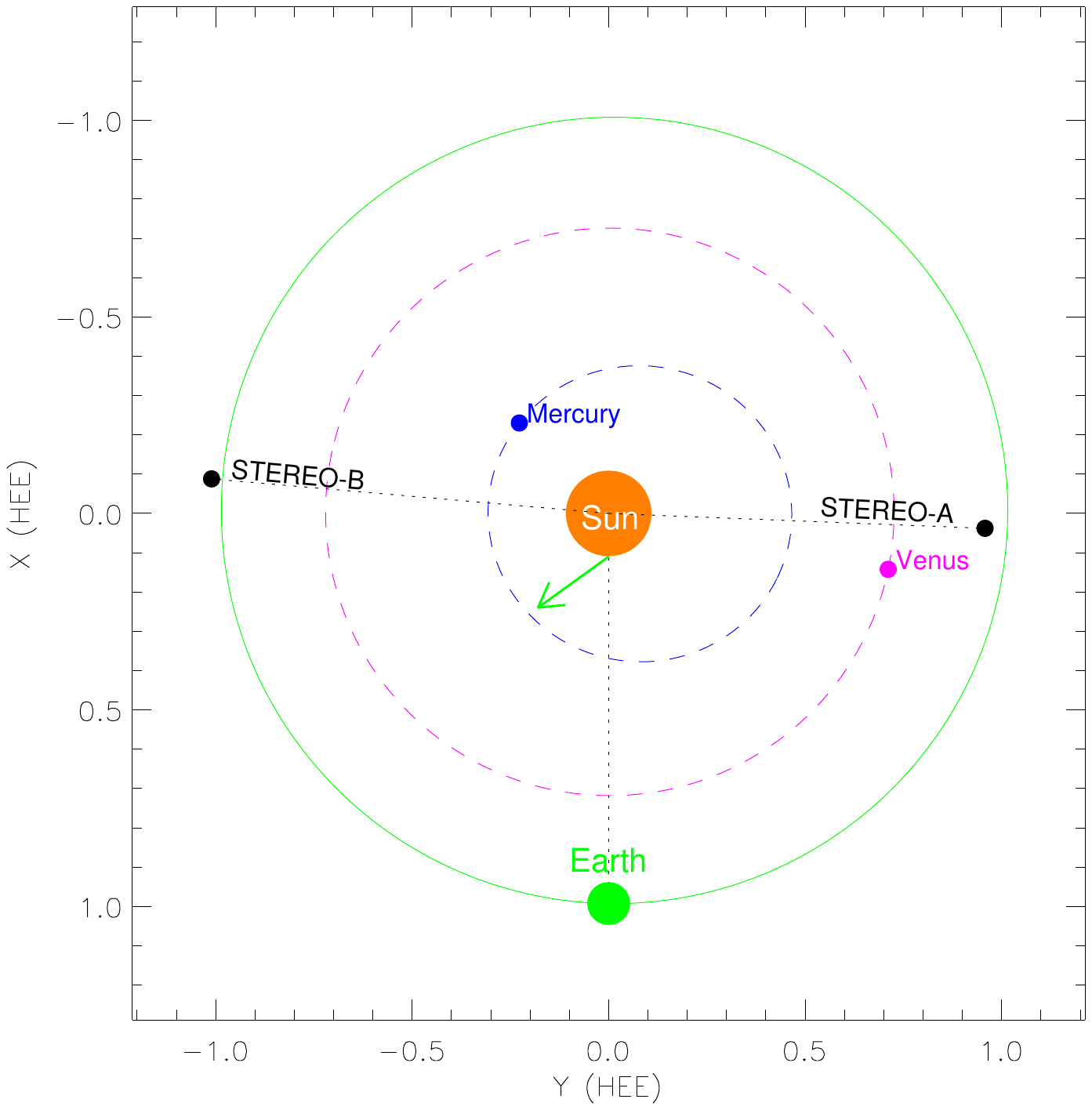} \caption{Positions of {\sl STEREO} two satellites
 relative to Sun (orange) and the orbit of Earth (green) in the x-y plane of the Heliocentric
 Earth Ecliptic coordinate system at 22:10 UT on March 09, 2011 (Units are in astronomical
 unit A. U.). The positions of Venus and Mercury are also marked. The green arrow points to
 the eruption direction of the blowout jet. The position of Earth, COR1 Ahead and COR1 Behind are perfect, and,
fortunately, have successfully imaged the blowout jet eruption.
\label{sat}}

\end{figure}

\begin{figure}[htbp]
\epsscale{1.} \plotone{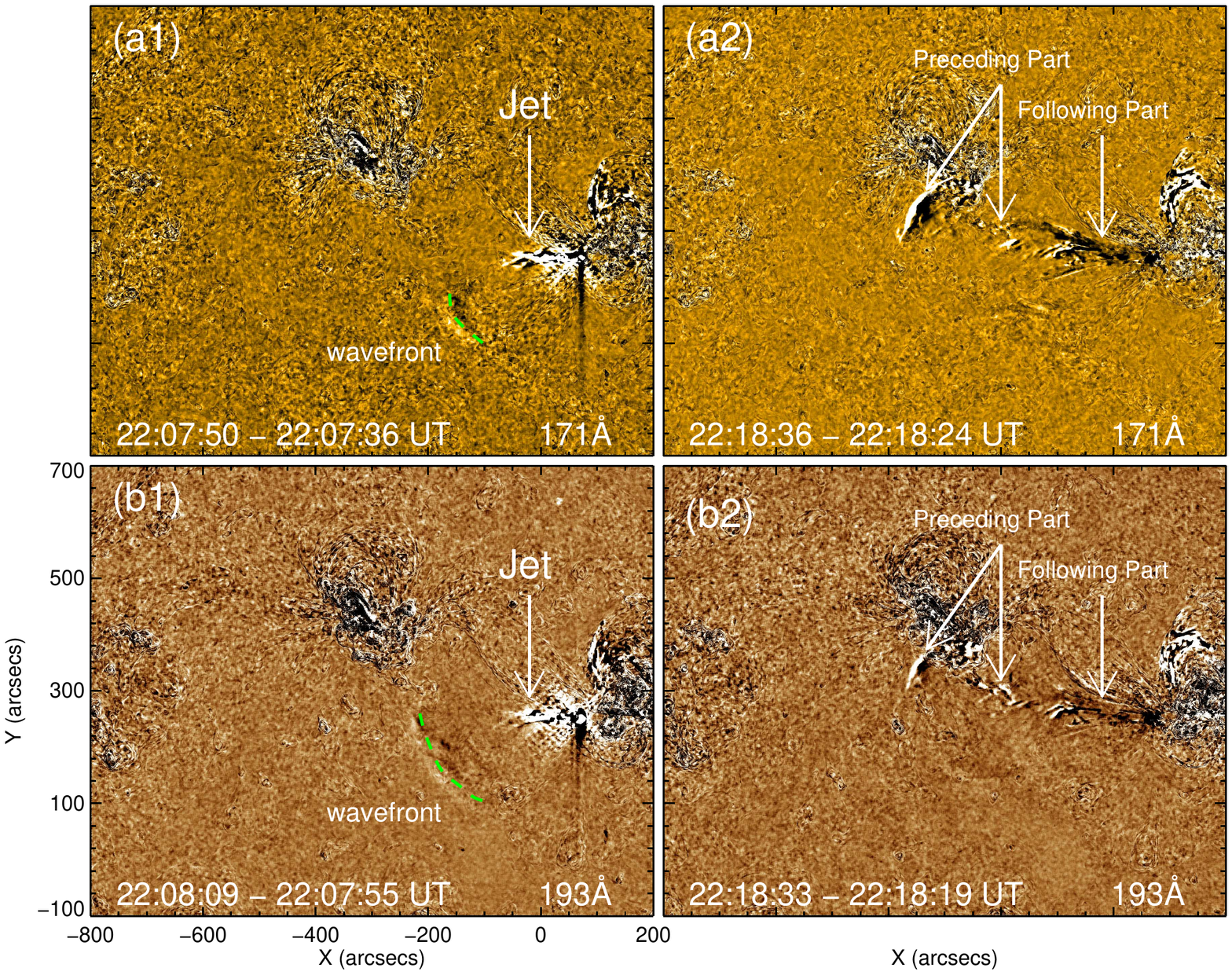}
\caption{{\sl SDO}/AIA 171 \AA\  and 193 \AA\ running difference images.
Panel (a1) and panel (b1) show the profile of the wavefront
with green dashed lines (see animation1.mpeg). Panels (a2) and (b2) show the preceding part
and the following part of the jet, respectively.
\label{wave}}

\end{figure}

\begin{figure}[htbp]
\plotone{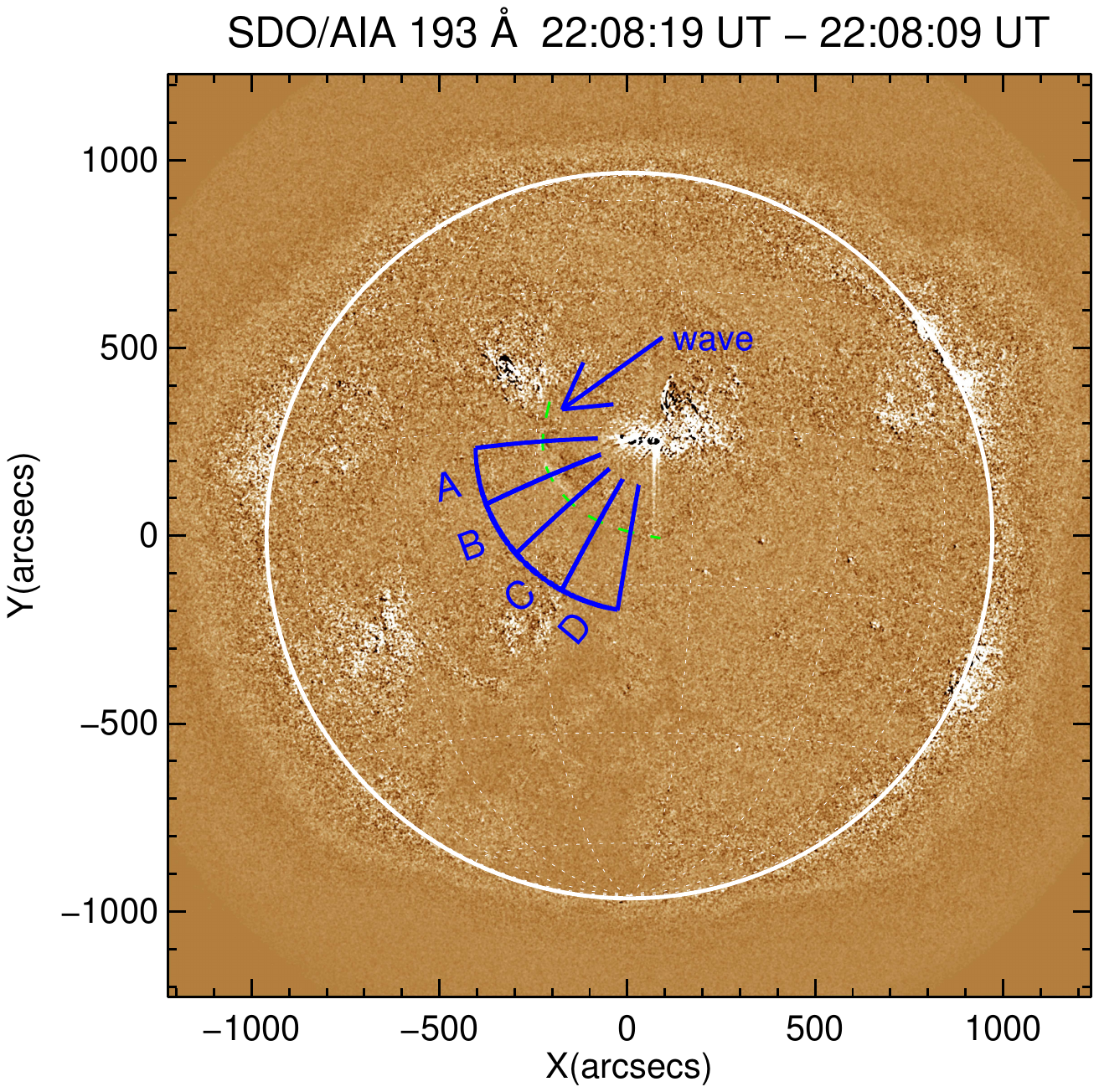} \caption{{\sl SDO}/AIA 193 \AA\ running difference
image showing four ${20}^{\circ}$ wide sectors (A to D), which are used to obtain the stacked
plots shown in Figure \ref{wave_velocity}. The green dashed line displays the wave profile.
\label{slice_wave_position}}
\end{figure}

\begin{figure}[htbp]
\plotone{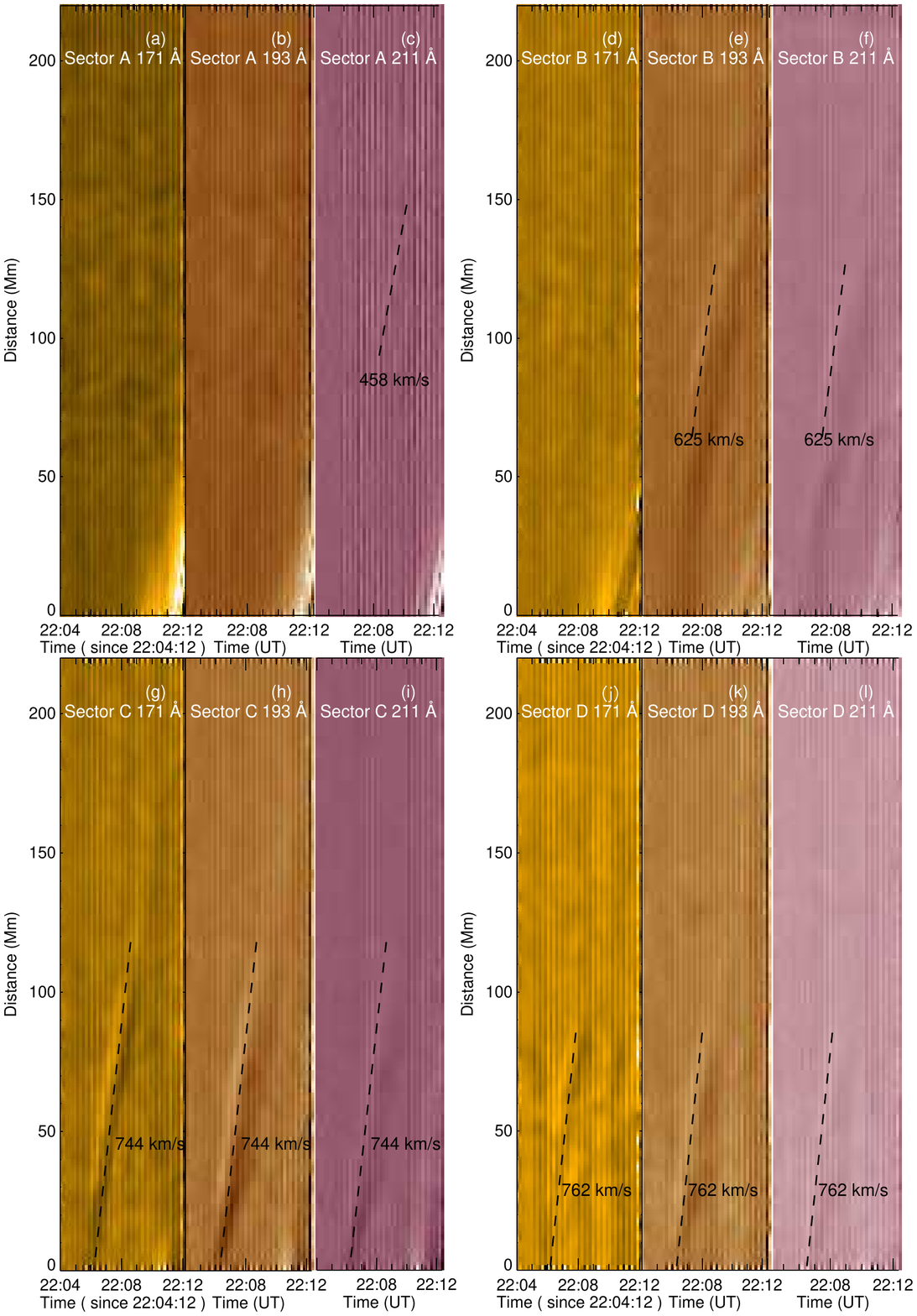}
\caption{Running difference stacked plots along Sector ``A'' to
Sector ``D'' in 171, 193, and 211 \AA. The black dashed lines
are displayed the wave velocity. The wave is too weak to show running
difference stacked plots at 171 \AA\ and 193 \AA\ along Sector ``A''.
\label{wave_velocity}}

\end{figure}

\begin{figure}[htbp]
\epsscale{1}
\plotone{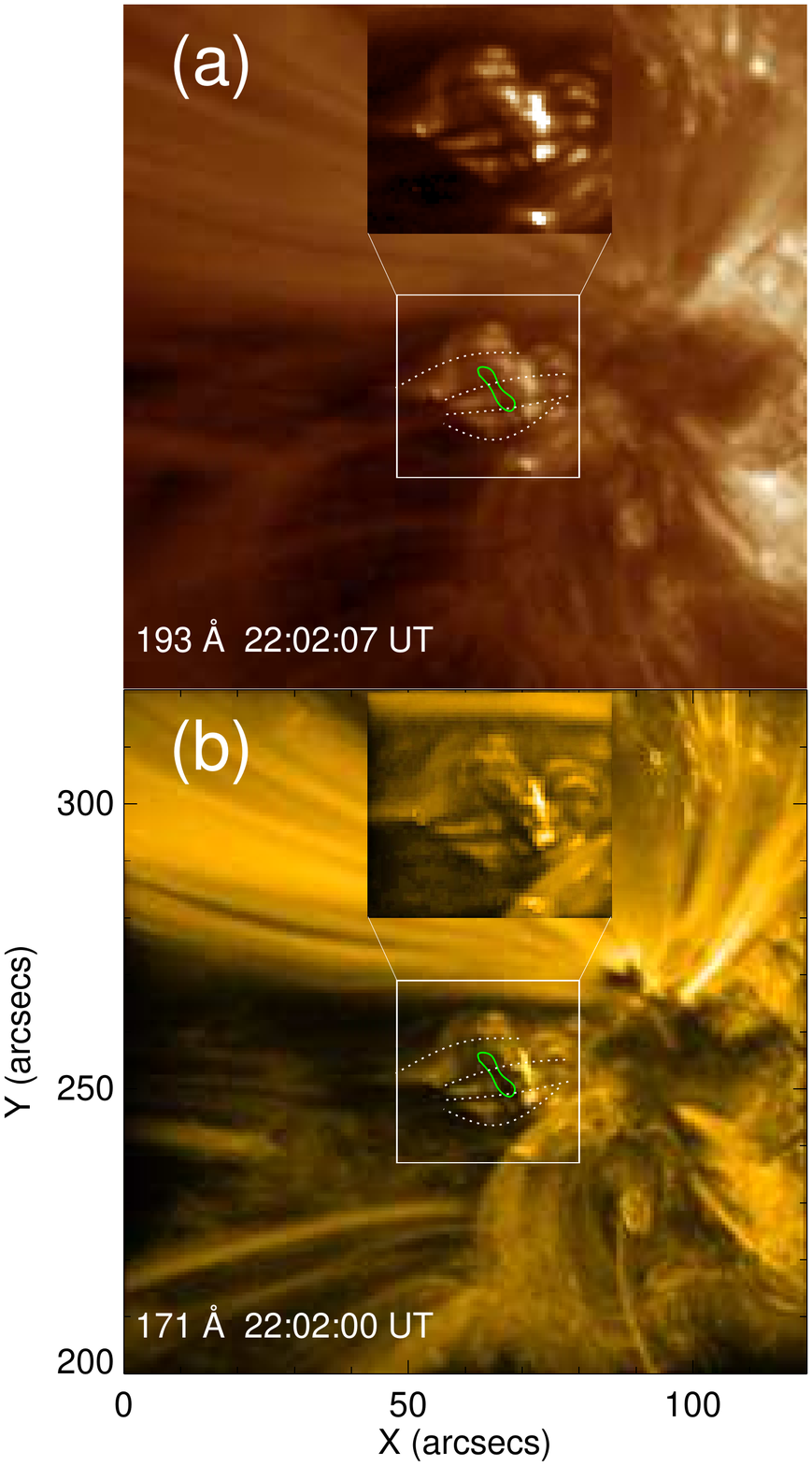}
\caption{Panels (a) and (b) are {\sl SDO}/AIA 193 and 171 \AA\ images, respectively.
The white boxes indicate the JBR. The white dotted lines
represent the small magnetic loops and the green contours display the profile of the mini-filament
located in panel (a) and panel (b). The inserted images show the the mini-filament and closed-loop system (see animation2.mpeg).}
 \label{filament}

\end{figure}

\begin{figure}[htbp]
\epsscale{1.0} \plotone{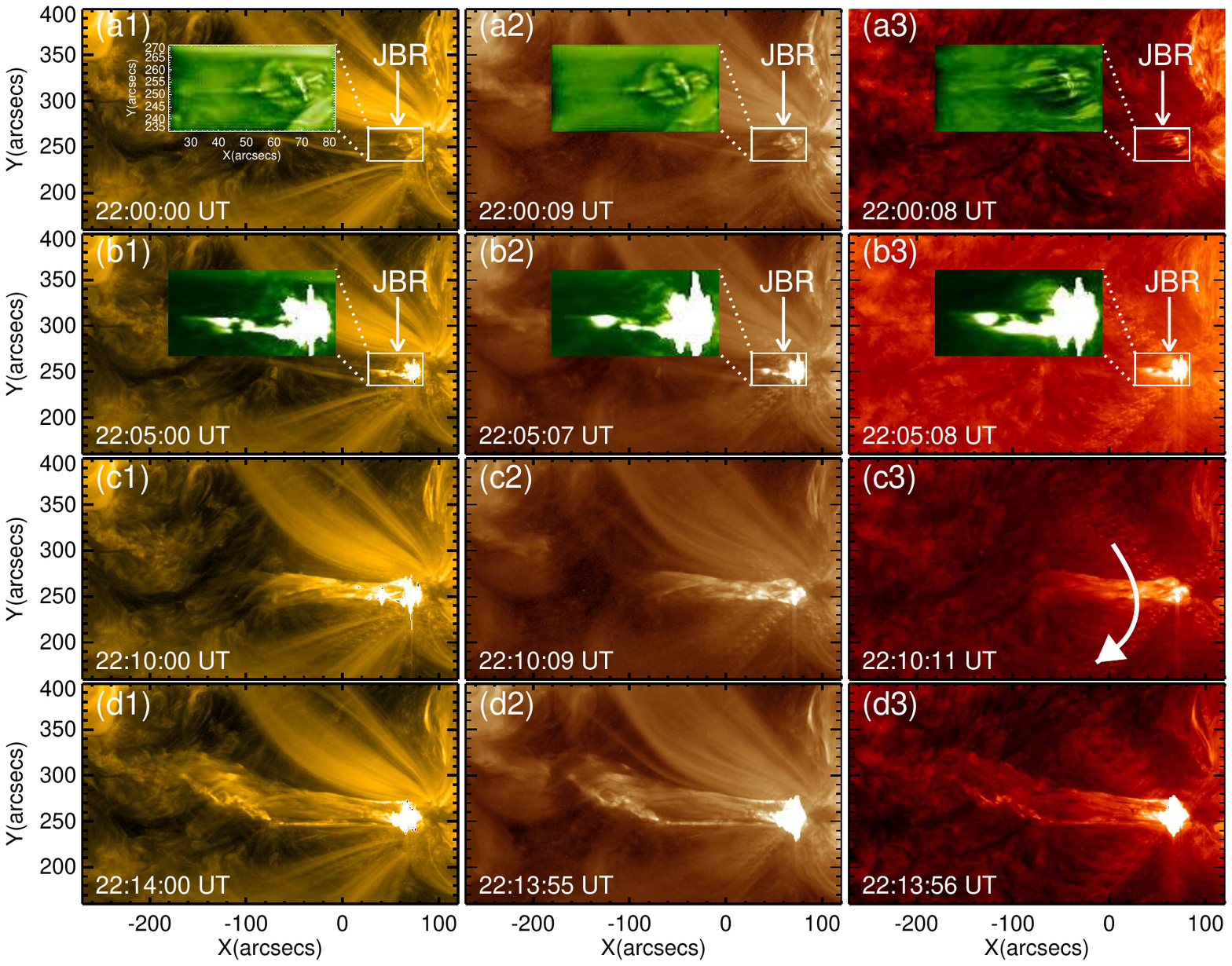}
\caption{{\sl SDO} multi-wavelength images at
171 \AA, 193 \AA, and 304 \AA\ highlighting the blowout jet eruption. The inserted images in the top two rows show the mini-filament location (panels from (a) to (b2)) of the JBR in the small rectangle region.
In panel (c2), a white curved arrow indicates the direction of untwisted motion. (see animation3.mpeg;
available in the accompanying online material).
\label{aia}}

\end{figure}

\begin{figure}[htbp]
\epsscale{1} \plotone {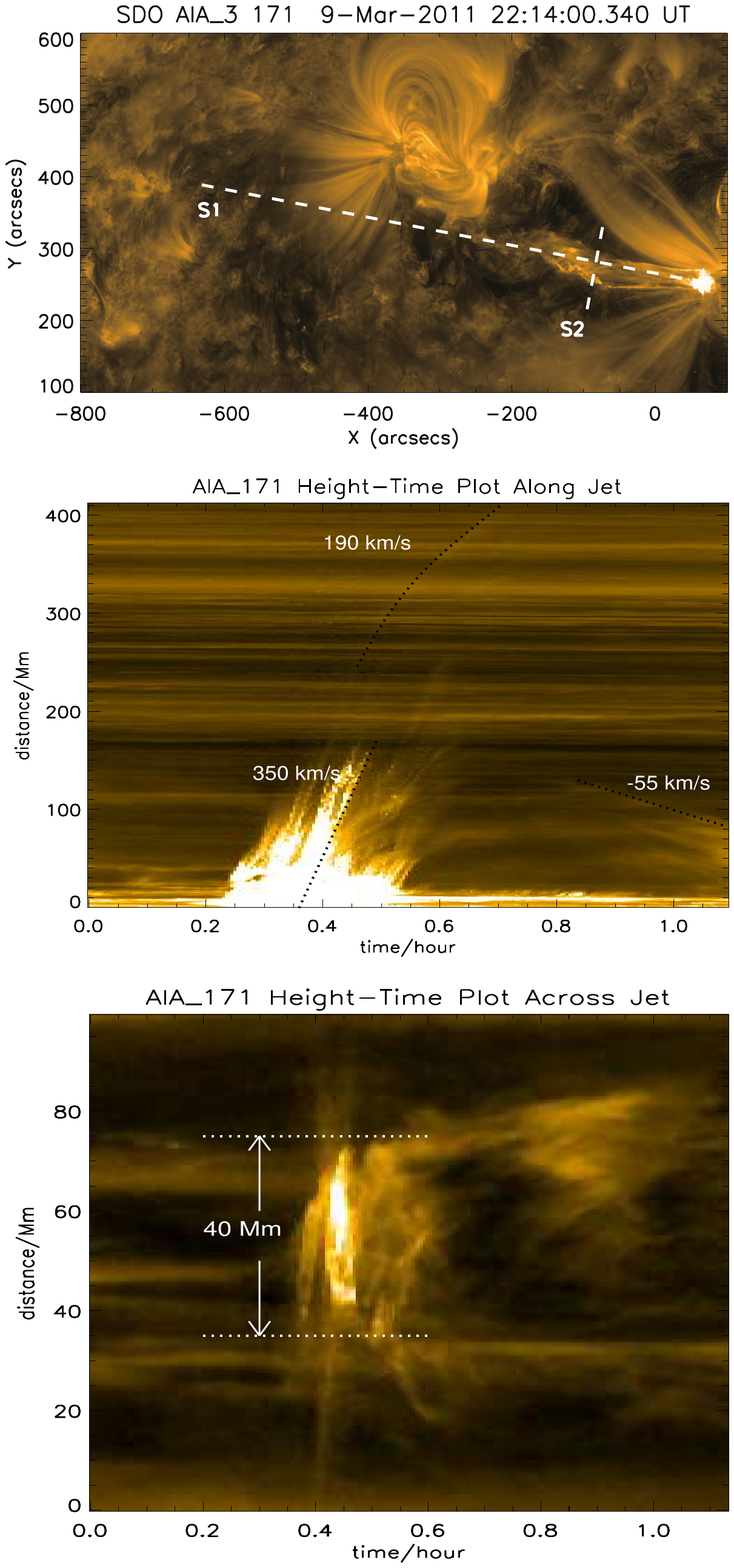}
\caption{Time-distance plot from the cross-cut along the jet with AIA 171 \AA\ image shown
in top image. The white dashed lines `` S1'' and `` S2'' indicate the trend along the jet and
vertical the jet, respectively. The main ejection speed of the jet is about \speed{350}, (bottom half),
while the following part velocity is about \speed{-55} (The two dotted lines are
linear fits, adopted in our computations). The jet preceding part average velocity is
about \speed{190} (the top part, a simple curvilinear fit is adopted in our computation). The result
of the height-time plots are reported in the bottom figure, where we measured the jet width (brighter area)
to be at least 40 Mm, indicating the very large jet scale.
\label{slice}}
\end{figure}

\begin{figure}
\epsscale{1}
\plotone{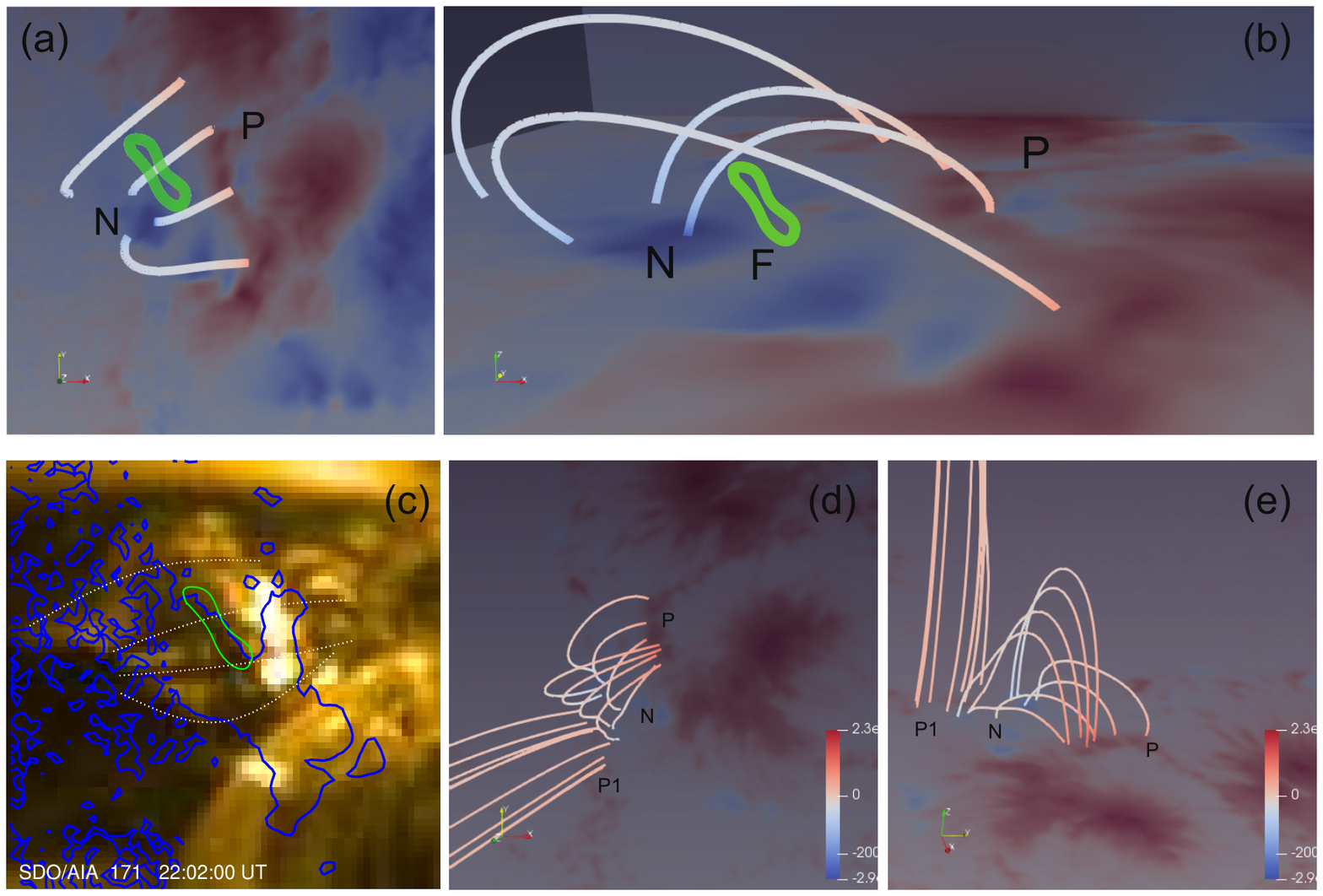}
\caption{The force-field extrapolation (FFE) model of the magnetic field structure
of the jet base. Panels (a) and (b) show the mini-filament (green contour line) and the
magnetic field structure above it. Panel (c) displays the mini-filament (green contour line) and the magnetic
neutral lines (blue lines) at 171 \AA. Panels (d) and (e) represent the locations of the ``P'', ``N'', and ``P1''.
The open magnetic field lines and close magnetic field lines are also shown in panels (d) and (e).
\label{extra}}
\end{figure}

\begin{figure}[htbp]
\epsscale{1.} \plotone {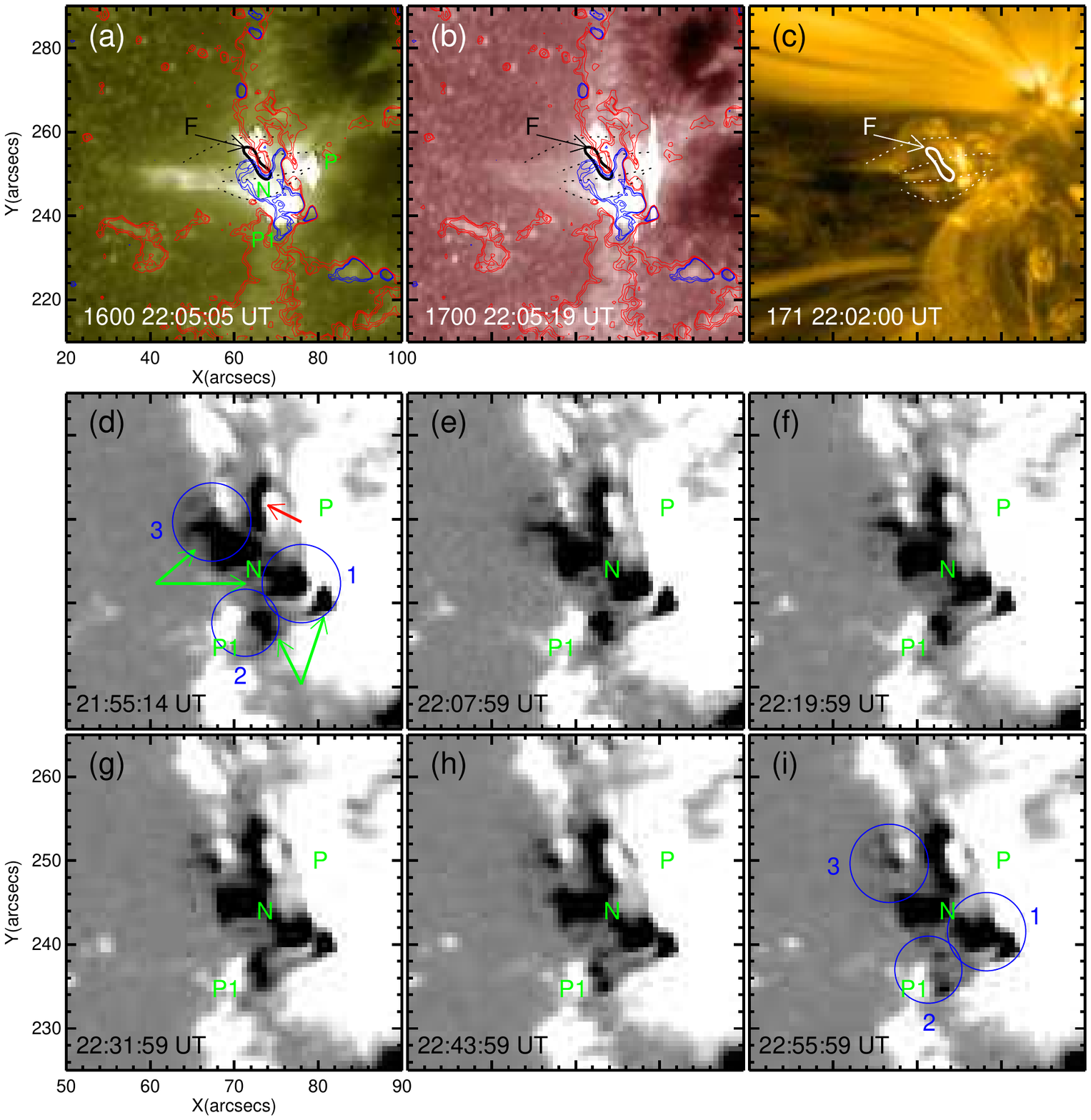}
\caption{{\sl SDO} 1600 \AA, 1700 \AA, 171 \AA, and LOS HMI images.
Panels (a) to (c) report the HMI image overlaying AIA images at 22:04:59 UT. The blue contours
represent the negative magnetic field and red contours represent the positive magnetic field.
The black contour dotted lines in panel (a), (b) and the white
contour dotted lines in panel (c) refer to the magnetic loops above the mini-filament. The mini-filament
profile are also shown in panels (a), (b), and (c), respectively. The contour levels are $\pm100$ G,
$\pm50$ G, $\pm30$ G. In panel (a), the green labels ``P'', ``N'', ``P1'' indicate the main
positive, emerging negative and the open positive magnetic field, respectively. From (d) to (i), a series
of LOS HMI images show the negative flux evolution of the jet base (see animation4.mpeg). The green arrows and red
arrow indicate the negative cancellation and emergence fluxes, respectively. In order to indicate the
detail of the cancellation, we mark three circles to display the very evident magnetic cancellation regions
(see labels ``1'', ``2'', and ``3'').
\label{hmi_euv}}
\end{figure}

\begin{figure}
\epsscale{1.} \plotone{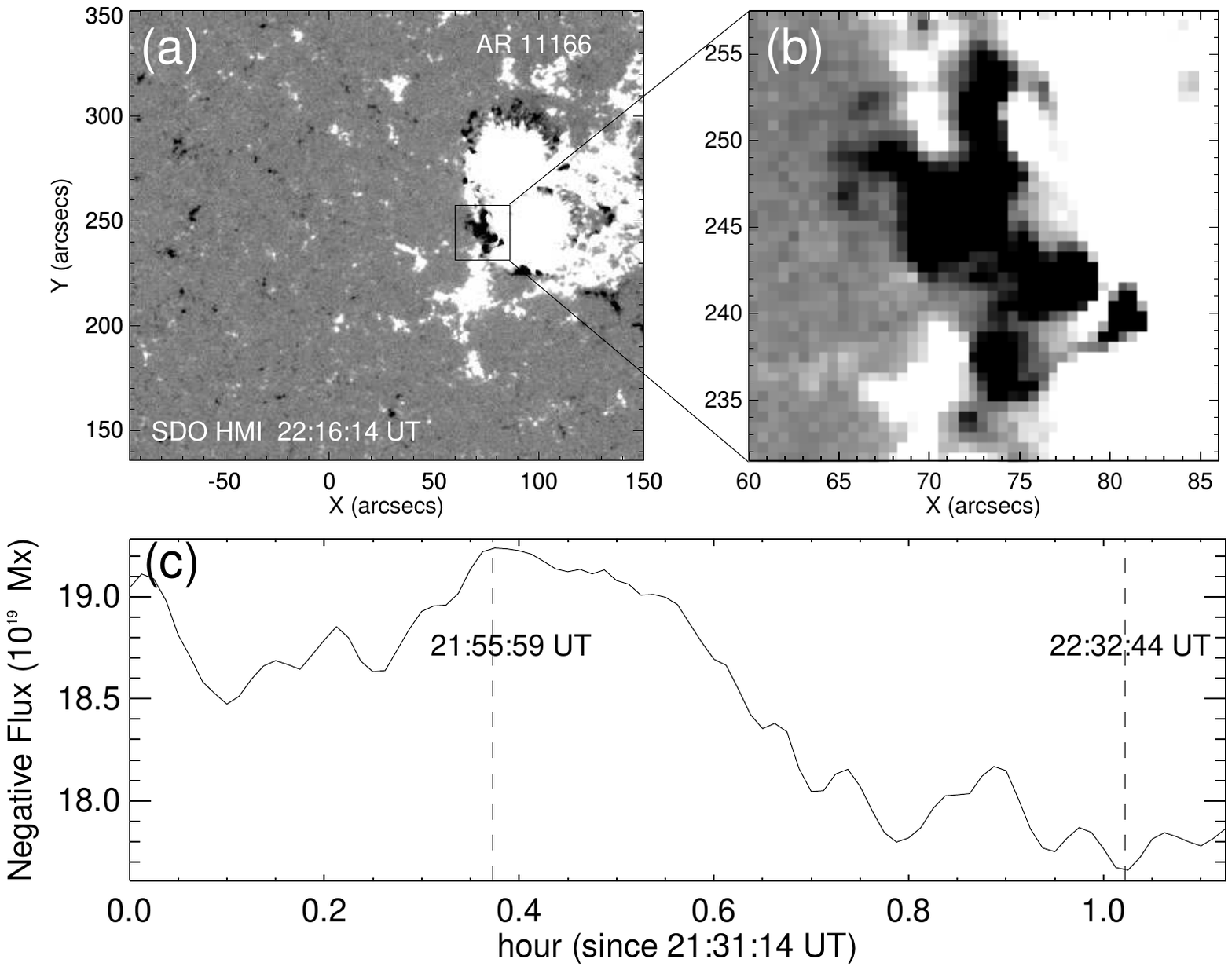}
\caption{Panel (a) shows the HMI LOS magnetogram with a small black
 box showing the position of the cancellation at the jet base. Panel (b) shows the zoomed view of the black box.
 Panels (c) shows the negative flux evolution in Panel (b) from 21:31:14 UT to 22:38:44 UT.
 \label{flux_all}}
\end{figure}



\begin{figure}
\epsscale{1.} \plotone{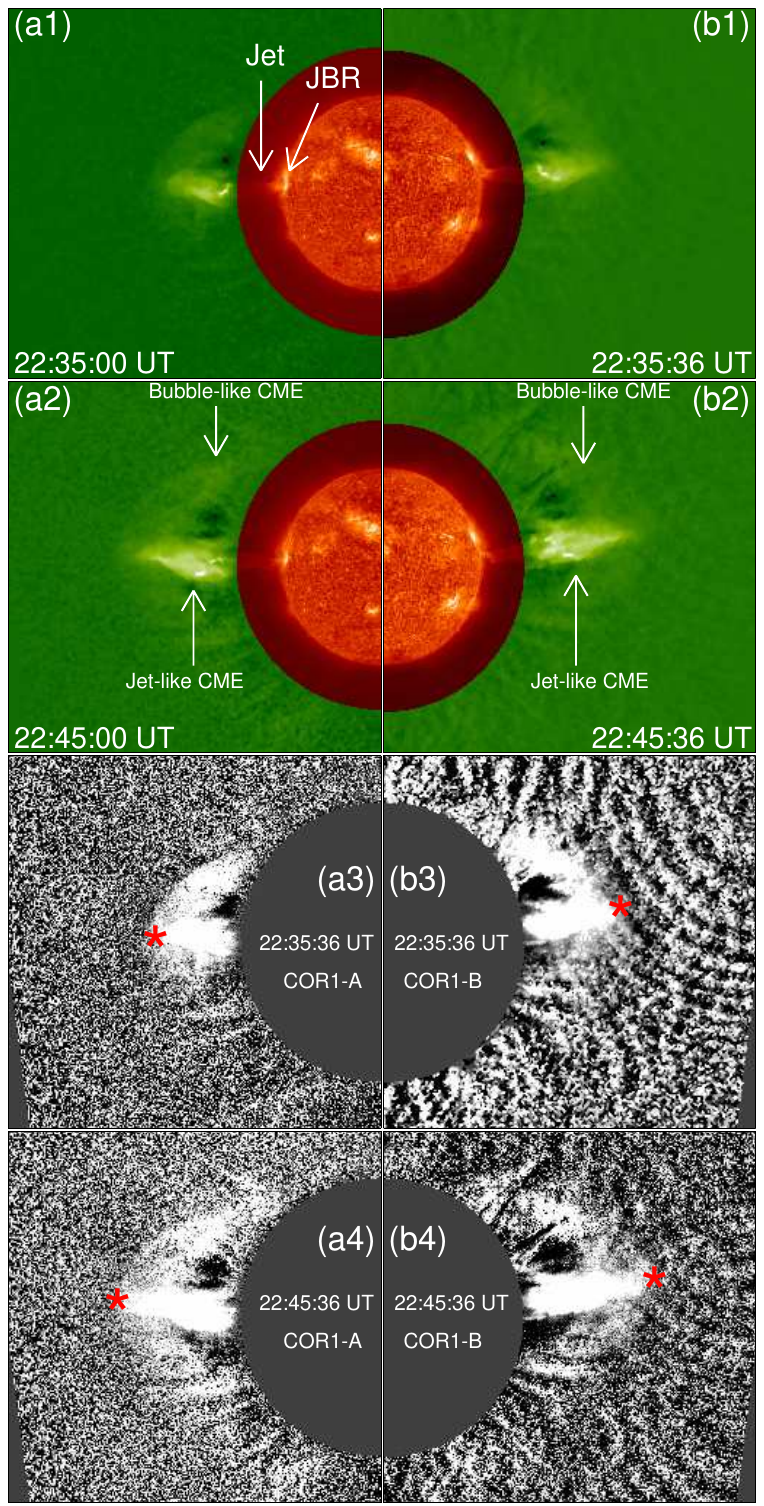}
\caption{Panels (a1), (a2) correspond to combing images of {\sl STEREO}
Ahead 304 \AA\ and COR1, and panels (b1), (b2) are {\sl STEREO} Behind 304 \AA\ and COR1 combination images,
respectively. The white arrows indicate the jet-like and bubble-like CME components. The third and the fourth columns show the COR1 Ahead and Behind running difference images of {\sl STEREO}. The
red ``$\ast$'' represents the top of the jet-like CME component in panels from (a3) to (b4). Note that the top of the jet-like CME component seems to exceed the front of the
bubble-like component in panels (b3) and (b4) (see animation5.mpeg).
\label{stereo-cme}}
\end{figure}

\end{document}